\documentclass[prd,showpacs,superscriptaddress,floatfix,notitlepage,nofootinbib,reprint]{revtex4-1}

\usepackage{graphicx}  
\usepackage{bm}        
\usepackage{amsmath,amssymb}   
\usepackage{color}
\usepackage{aas_macros}
\usepackage{color}
\usepackage{url}
\usepackage{hyperref}

\newcommand{\mr}{\mathrm}
\newcommand{\mb}{\mathbf}
\renewcommand{\d}{\mathrm{d}}

\newcommand{\bea}{\begin{eqnarray}}
\newcommand{\eea}{\end{eqnarray}}
\newcommand{\be}{\begin{equation}}
\newcommand{\ee}{\end{equation}}

\def\Mpc{{\rm Mpc}}

\begin{document}
\widetext

\title{Cosmic tidal reconstruction}

\author{Hong-Ming Zhu}
\affiliation{Key Laboratory for Computational Astrophysics,
National Astronomical Observatories, Chinese Academy of Sciences,
20A Datun Road, Beijing 100012, China}
\affiliation{University of Chinese Academy of Sciences, Beijing 100049, China}

\author{Ue-Li Pen}
\affiliation{Canadian Institute for Theoretical Astrophysics, 60 St. George Street, Toronto, Ontario M5S 3H8, Canada}
\affiliation{Canadian Institute for Advanced Research, CIFAR Program in Gravitation and Cosmology, Toronto, Ontario M5G 1Z8, Canada}
\affiliation{Perimeter Institute for Theoretical Physics, 31 Caroline Street North, Waterloo, Ontario, N2L 2Y5, Canada}

\author{Yu Yu}
\affiliation{Key Laboratory for Research in Galaxies and Cosmology,
Shanghai Astronomical Observatory, Chinese Academy of Sciences,
80 Nandan Road, Shanghai 200030, China}

\author{Xinzhong Er}
\affiliation{Key Laboratory for Computational Astrophysics,
National Astronomical Observatories, Chinese Academy of Sciences,
20A Datun Road, Beijing 100012, China}
\affiliation{INAF, Osservatorio Astronomico di Roma, via Frascati 33, 
I-00040, Monteporzio Catone, Italy}

\author{Xuelei Chen}
\affiliation{Key Laboratory for Computational Astrophysics,
National Astronomical Observatories, Chinese Academy of Sciences,
20A Datun Road, Beijing 100012, China}
\affiliation{University of Chinese Academy of Sciences, Beijing 100049, China}
\affiliation{Center of High Energy Physics, Peking University, Beijing 100871, China}

\date{\today}

\begin{abstract}
The gravitational coupling of a long-wavelength tidal field with small-scale 
density fluctuations leads to anisotropic distortions of the locally measured 
small-scale matter correlation function.
Since the local correlation function is known to be statistically isotropic in 
the absence of such tidal interactions, the tidal distortions can be used to 
reconstruct the long-wavelength tidal field and large-scale density field in
analogy with the cosmic microwave background lensing reconstruction. 
In this paper we present the theoretical framework of cosmic tidal 
reconstruction and test the reconstruction in numerical simulations.
We find that the density field on large scales can be reconstructed with good 
accuracy and the cross-correlation coefficient between the reconstructed 
density field and the original density field is greater than 0.9 on large scales
($k\lesssim0.1\ h/\mr{Mpc}$), with the filter scale $\sim1.25\ \mr{Mpc}/h$. 
This is useful in the 21cm intensity mapping survey, where the long-wavelength 
radial modes are lost due to a foreground subtraction process. 
\end{abstract}

\maketitle

\section{Introduction}
The large-scale structure contains a wealth of information about our Universe.
The cosmic acceleration, neutrino masses, early Universe models and other 
properties of the Universe can  be inferred with current or upcoming surveys.
The initially linear and Gaussian density perturbations grow due to 
gravitational interaction; the nonlinearities develop and induce the couplings 
between different modes.
This leads to striking non-Gaussian features in the large-scale structure
of the Universe, limiting the cosmological information that can be extracted 
from galaxy surveys.
However, such correlations between the different perturbation modes on different
scales imply that we could infer the large-scale density field by observing 
small-scale density fluctuations.

The basic idea is that the evolution of small-scale density perturbations 
$\delta(\bm{k}_S)$ is modulated by long-wavelength perturbations $\Phi_L$, 
causing both isotropic and anisotropic distortions of the local small-scale 
power spectrum \cite{2014:tidal}. 
The isotropic distortion, which only depends on the magnitude of the wave vector
$\bm{k}_S$, is mainly due to the change in the background density from the 
long-wavelength density perturbation $\nabla^2\Phi_{L}\propto\delta_L$, where 
$\Phi_L$ denotes the long-wavelength gravitational potential.
The anisotropic distortion, which also depends on the direction of $\bm{k}_S$,
is induced by the long-wavelength tidal field 
$t_{ij}=\Phi_{L,ij}-\delta_{ij}\nabla^2\Phi_{L}/3$, where 
$\Phi_{L,ij}\equiv \partial^2\Phi_L/\partial x^i \partial x^j$.
Using the isotropic modulation on the small-scale power spectrum to reconstruct
the large-scale density field has been studied in Ref. \cite{2014li}.
However, the anisotropic distortions are more robust since many other processes
can lead to isotropic distortions in the local small-scale power spectrum. 
By applying quadratic estimators, which quantify the local anisotropy of 
small-scale density statistics, the long-wavelength tidal field can be 
reconstructed accurately \cite{2012:pen}.
The reconstruction of the long-wavelength tidal field from the local small-scale
power spectrum is similar to the reconstruction of shear fields in gravitational
lensing \cite{2008:lu,2010lu}.

In this paper, we investigate the anisotropic method in detail. 
The evolution of small-scale density fluctuations in the presence of the 
long-wavelength tidal field $t_{ij}$ has been studied extensively in 
Ref. \cite{2014:tidal}.
The tidal distortion of the local small-scale power spectrum is given by 
\begin{equation}
P(\bm{k}_S,\tau)|_{t_{ij}}=P_{1s}(k_S,\tau)+
\hat{k}_S^i\hat{k}_S^jt_{ij}^{(0)}P_{1s}(k_S,\tau)f(k_S, k_L,\tau),
\end{equation}
where $\hat{\bm{k}}$ denotes the unit vector in the direction of $\bm{k}$,
$P_{1s}(k_S,\tau)$ is the isotropic linear power spectrum, $f(k_S, k_L, \tau)$ 
describes the coupling of the long-wavelength tidal field to the small-scale 
density fluctuations and superscript $(0)$ denotes some ``initial'' time. 
The traceless $3\times3$ tensor $t_{ij}$ can be decomposed into five 
independently observable components 
\begin{eqnarray}
t_{ij}=\left( \begin{array}{ccc}
\gamma_{1}-\gamma_{z} & \gamma_{2} & \gamma_{x}\\
\gamma_{2} & -\gamma_{1}-\gamma_{z} & \gamma_{y}\\
\gamma_{x} & \gamma_{y} & 2\gamma_z
\end{array} \right),
\end{eqnarray}
where $\gamma_{1}=(\Phi_{L,11}-\Phi_{L,22})/2$ and $\gamma_{2}=\Phi_{L,12}$
are the only two tidal shear fields that do not involve $i=3$
(see Sec. \ref{sec:3b} for more details).
In this paper we take the third axis to be parallel with the line of sight 
$(x_\parallel=x^3)$ and the first and second axes to lie in the tangential
plane perpendicular to the line of sight $[\bm{x}_\perp=(x^1,x^2)]$.
The anisotropic distortion can also be decomposed into five orthogonal 
quadrupolar distortions as in weak lensing.
While $t_{ij}$ in principle has five independently observable components,
in this paper we shall focus on the two transverse shear terms,
$(\hat{k}_1^2-\hat{k}_2^2)\gamma_1^{(0)}P_{1s}f$
and $2\hat{k}_1\hat{k}_2\gamma_2^{(0)}P_{1s}f$, which describe quadrupolar 
distortions in the tangential plane perpendicular to the line of sight.  
The quadrupolar distortions containing $i=3$ will be affected by peculiar 
velocities and require additional treatments beyond the scope of this paper. 
These two tidal shear fields $\gamma_1$ and $\gamma_2$ can be converted into 
the two-dimensional (2D) convergence field,
$\kappa_\mr{2D}\equiv (\Phi_{L,11}+\Phi_{L,22})/2$, using the relation 
\begin{equation}
\kappa_{\mr{2D},11}+\kappa_{\mr{2D},22}=\gamma_{1,11}-\gamma_{1,22}+2\gamma_{2,12}.
\end{equation}
We can obtain the three-dimensional (3D) convergence field 
$\kappa_\mr{3D}\equiv\nabla^2\Phi_L/3$ from the 2D convergence by 
\begin{equation}
\kappa_\mr{3D,11}+\kappa_\mr{3D,22}=\frac{2}{3}\nabla^2\kappa_\mr{2D}.
\end{equation} 
The large-scale density field $\delta_L$ is then given by the 3D convergence 
field through the Poisson equation.
The transverse shear fields are evaluated by applying quadratic estimators 
to the density field as in weak lensing.
The optimal tidal shear estimators can be derived under the Gaussian assumption.

We call this method {\it cosmic tidal reconstruction}, as it exploits the
local anisotropic features arising from tidal interactions.
Each independent measurement of the small-scale power spectrum gives some 
information about the power spectrum on large scales. 
Thus, instead of being limited by the non-Gaussianity on small scales, we exploit these strong correlations to improve the measurement of the large-scale 
structures. 

Since we only use the tidal shear fields in the $\bm{x}_\perp$ plane to
reconstruct the long-wavelength density field, the changes of the 
long-wavelength density field $\delta_L$ along the $x_\parallel$ axis are
inferred from the variations of tidal shear fields $\gamma_1$ and $\gamma_2$ 
along this direction, i.e., we reconstruct these modes ``indirectly.''
We cannot capture rapid changes of the density field along the $x_\parallel$ 
axis, i.e., those modes with large $k_\parallel$. This is reflected by the 
anisotropic noise of the reconstructed mode $\kappa_\mr{3D}(\bm{k})$, which is 
$\sigma^2_{\kappa_\mr{3D}}\propto (k^2/k_\perp^2)^2$ (see Sec. \ref{sec:3b} for 
derivations).  
The noise is infinite when $k_\perp\to0$. 
In this case, we expect these modes contain nothing but noise. 
The well-reconstructed modes are those with small $k_\parallel$ and large
$k_\perp$. 
This is useful in the cross correlations of the 21cm intensity 
mapping survey with other cosmic probes, such as integrated Sachs-Wolf effect, lensing, 
photo-$z$ galaxies, optical depth,
and others \cite{2007ApJ...660.1030F,2008MNRAS.384..291A}. 
These low $k_\parallel$ modes are normally lost in the 21cm intensity mapping data 
due to foreground subtraction \cite{2012ApJ...752..137M}.

The local quadrupolar distortions have also been discussed in
Refs. \cite{masui2010,2012:jeong} where the theoretical constructions were 
outlined.
The basic idea of purely transverse tidal reconstruction has been studied
in Ref. \cite{2012:pen}. This paper expands on the early work by considering 
the tidal reconstruction method in detail. This paper is organized as follows.
In Sec. \ref{sec:deform}, we present the basic formalism.
In Sec. \ref{sec:recon}, we construct the tidal shear estimators and 
present the reconstruction algorithm. 
In Sec. \ref{sec:sim}, we study the tidal reconstruction in $N$-body 
simulations. 
In Sec. \ref{sec:disc}, we examine the validity of cosmic tidal 
reconstruction and discuss its future applications.

\section{Tidal deformation}
\label{sec:deform}
We first present a heuristic and intuitive description of the tidal interaction
before going through the detail calculations.
The motion of a dark matter fluid element in the Universe includes translation,
rotation, and deformation. 
Let us consider two neighboring points $P_0$ at $\bm{x}_0$ 
and $P$ at $\bm{x}=\bm{x}_0+\Delta\bm{x}$ in this fluid element at time $t_0$,
with velocities
$\bm{v}|_{P_0}$ and $\bm{v}|_P$, respectively.   
Here, $\bm{v}|_{P}$ can be expanded as
\begin{eqnarray}
v^i|_{P}=v^i|_{P_0}+v^i_{\ ,j}|_{P_0}\Delta x^j,
\end{eqnarray}
where $v^i_{\ ,j}=\partial v^i/\partial x^j$ is the velocity gradient tensor.
It can be decomposed into a symmetric part $S^i_{\ j}=(v^i_{\ ,j}+v_j^{\ ,i})/2$
and an antisymmetric part $A^i_{\ j}=(v^i_{\ ,j}-v_j^{\ ,i})/2$. The velocity of
$P$ now becomes
\begin{eqnarray}
v^i|_{P}=v^i|_{P_0}+(S^i_{\ j}+A^i_{\ j})|_{P_0}\Delta x^j.
\end{eqnarray}
The tensor $A^i_{\ j}$ describes the rotation of this 
fluid element.
The symmetric tensor $S^i_{\ j}$ is called the 
deformation velocity tensor in fluid mechanics, with diagonal components 
describing the stretching along three axes and off-diagonal components 
describing the shear motion. 

In the free-falling reference system of this fluid element, $v|_{P_0}=0$, with
$P_0$ as the origin point.
Here, we focus on the deformation, neglecting the rotation.
Then, the position of $P$ at time $t$ is given by 
\begin{eqnarray}
x^i(t)|_P=x^i+\int_{t_0}^tS^i_{\ j}(t{'})|_{P_0}dt{'}x^j,
\end{eqnarray}
where $x^i=\Delta x^i$ is the initial separation between $P_0$ and $P$.
The shape tensor $\mathfrak{S}^i_{\ j}=\int_{t_0}^tS^i_{\ j}(t{'})|_{P_0}dt{'}$ 
describes the deformation. 
In the local inertial frame, the dark matter fluid element
is only sensitive to the residual anisotropy of gravitational forces,
which causes the fluid element to deform. 
It is the same as ocean tides on the Earth induced by the tidal forces from 
the Moon and the Sun.
We call such effects induced by the surrounding large-scale structure {\it cosmic tides} \cite{2012:pen} in analogy to ocean tides on the Earth.  
By observing these local anisotropies, we can measure the large-scale tidal
field and density field.
In reality, the interaction between a long-wavelength tidal field
and small-scale density fluctuations is more complicated than the displacement 
of particles in a fluid element, we discuss this in more detail below. 

The effect of a general long-wavelength tidal field on the evolution of 
small-scale density perturbations with $k_L\ll k_S$
has been studied in Ref. \cite{2014:tidal}. 
Here, we focus on the effect of the traceless tidal field 
$t_{ij}=\Phi_{L,ij}-\delta_{ij}\nabla^2\Phi_L/3$ from a long-wavelength scalar
perturbation $\Phi_L$.

In the expanding Universe, the equation of motion for a particle is 
\begin{equation}
\label{eq1}
\frac{d^2{\bm x}}{d\tau}+\mathcal{H}(\tau)\frac{d{\bm x}}{d\tau}
=-\nabla_{\bm x}\phi,
\end{equation}
where $\bm{x}$ is the comoving Eulerian coordinate, $\tau$ is the conformal 
time, $\mathcal{H}(\tau)=d\mr{ln}a/d\tau$ is the comoving Hubble parameter, 
$a(\tau)$ is the scale factor and $\phi$ is the gravitational potential.
In Lagrangian perturbation theory, the dynamical variable is the Lagrangian 
displacement field $\mathbf{s}(\bm{q},\tau)$, defined as
\begin{equation}
\label{eq2}
\bm{x}(\bm{q},\tau)=\bm{q}+\mb{s}(\bm{q},\tau),
\end{equation}
where $\bm{q}$ is the comoving Lagrangian coordinate. The displacement field
maps the initial particle position $\bm{q}$ into the final Eulerian position
$\bm{x}$.
The density contrast $\delta(\bm{x})$ is given by the mass conservation relation,
\begin{eqnarray}
\label{eq3}
\delta(\bm{x}(\bm{q},\tau))=\frac{1}{J(\bm{q},\tau)}-1,
\end{eqnarray}
where $J(\bm{q},\tau)=\mr{det}(\delta^i_{\ j}+\mr{M}^i_{\ j}(\bm{q},\tau))$
and $\mr{M}^i_{\ j}=\partial\mr{s}^i/\partial q^j$.
We also have 
$\partial/\partial q^i=\partial/\partial x^i+\mr{M}^j_{\ i}\partial/\partial x^j$.

Now we want to study the evolution of small-scale density perturbations  
in the presence of the long-wavelength tidal field, so the gravitational 
potential $\phi$ in Eq. (\ref{eq1}), which drives the motion of a particle, 
contains not only the part sourced by the small-scale density 
fluctuations, $\Phi_s$, but also  a part induced by the long-wavelength tidal 
field \cite{2014:tidal},
\begin{eqnarray}
\label{eq4}
\phi(\bm{x}(\bm{q},\tau))=\Phi_s(\bm{x}(\bm{q},\tau))
+\frac{1}{2}t_{ij}(\bm{0}, \tau)x^i x^j.
\end{eqnarray}
The tidal field can be written as $t_{ij}(\bm{0}, \tau)=T(\tau)t_{ij}^{(0)}(\bm{0})$, 
where superscript  $(0)$ denotes the tidal field evaluated at the ``initial'' time $\tau_0$,
$T(\tau)=D(\tau)/a(\tau)$ is the linear transfer function, 
$D(\tau)$ is the linear growth function, and $D(\tau_0)=a(\tau_0)=1$.
Note that at the origin of the free-falling frame the contribution from the 
tidal field is zero. The small-scale potential $\Phi_s$ satisfies the Poisson 
equation,
\begin{eqnarray}
\label{eq5}
\nabla^2\Phi_s=4\pi Ga^2\bar{\rho}\delta_s=
\frac{3}{2}\Omega_m(\tau)\mathcal{H}^2\delta_s,
\end{eqnarray}
where $\Omega_m(\tau)$ is the density parameter at $\tau$.

The above equations can be solved perturbatively.
First, we decompose the displacement as
\begin{eqnarray}
\mb{s}=\mb{s}_s+\mb{s}_t,
\end{eqnarray}
where $\mb{s}_s$ is the contribution arising from the small-scale potential 
and $\mb{s}_t$ is due to the long-wavelength tidal field. Then 
$\mr{M}^i_{\ j}$ can also be decomposed into
$\mr{M}^{\ i}_{s\ j}=\partial \mr{s}_s^i/\partial q^j$ and
$\mr{M}^{\ i}_{t\ j}=\partial \mr{s}_t^i/\partial q^j$. 
Here, we only consider
the linear displacement $\mb{s}_{1s}$, $\mb{s}_{1t}$, and the quadratic term 
$\mb{s}_{2t}$ from the coupling of $\mb{s}_{1s}$ and $\mb{s}_{1t}$, and neglect
the terms of order $(\mb{s}_{1s})^2$,
which are nonlinear interactions between small-scale modes.
In the follow calculations, we focus on the coupling between the long mode 
(large-scale tidal field) and the short mode (small-scale density field), 
assuming that both large and small scale density fields follow 
linear evolutions.

At linear order, Eq. (\ref{eq1}) becomes two equations for 
$\mb{s}_{1s}$ and $\mb{s}_{1t}$, respectively,
\begin{eqnarray}
\label{eq:s1s}
\left[\frac{d^2}{d\tau^2}+\mathcal{H}\frac{d}{d\tau}\right]\mr{s}_{1s}^i(\bm{q},\tau)&=&
-{\partial}_q^i\Phi_{1s}(\bm{q},\tau),\\
\label{eq:s1t}
\left[\frac{d^2}{d\tau^2}+
\mathcal{H}\frac{d}{d\tau}\right]\mr{s}_{1t}^i(\bm{q},\tau)&=&
-\frac{1}{2}{\partial}^i_q[t_{kl}(\tau)q^kq^l], 
\end{eqnarray}
where $\partial_q^i =\delta^{ij}\partial/\partial q^j$, 
$\nabla_q^2\Phi_{1s}=3\Omega_m(\tau)\mathcal{H}^2\delta_{1s}/2$ and
$\delta_{1s}=-\mr{s}_{1s,i}^{\ i}$.  Eq. (\ref{eq:s1s}) can be solved to get 
\begin{eqnarray}
\mr{s}_{1s}^i(\bm{q},\tau)
=-\frac{\partial^i_q}{\nabla^2_q}\delta_{1s}(\bm{q},\tau)
=-D(\tau)\frac{\partial^i_q}{\nabla^2_q}\delta_{1s}(\bm{q},\tau_0).
\end{eqnarray}
The operator ${1}/{\nabla^2}$ denotes the inverse operator for $\nabla^2$. 
Equation (\ref{eq:s1t}) describes the evolution of the displacement induced 
by the long-wavelength tidal field and can be integrated to get 
\begin{eqnarray}
\mr{s}_{1t}^i(\bm{q},\tau)=-F(\tau)t^{(0)i}_{\ \ \ \ j}q^j,
\end{eqnarray}
where 
$F(\tau)=\int_0^\tau{d\tau{''}}{a(\tau{''})}T(\tau'')G(\tau-\tau'')$
and $G(\tau-\tau'')=\int^\tau_{\tau''}d\tau'/a(\tau')$.
The induced linear density fluctuation $\delta_{1t}$ is given by
\begin{eqnarray}
\delta_{1t} = -\mr{s}^{\ i}_{1t,i} = F(\tau)t^{(0)i}_{\ \ \ \ i}.
\end{eqnarray}
The trace of the tidal field $t_{ij}$ is zero, i.e. $\delta_{1t}=0$, 
so there is no first-order contribution to the density from the tidal field.

The evolution equation for $\mb{s}_{2t}$ involves quadratic mixed terms from
the coupling between $\mb{s}_{1s}$ and $\mb{s}_{1t}$. Inserting the Poisson 
equation to Eq. (\ref{eq1}) and subtracting the evolution equations for $\mb{s}_{1s}$,
$\mb{s}_{1t}$ and $\mb{s}_{2s}$ leads to
\begin{eqnarray}
\label{eq18}
&&\frac{d^2}{d\tau^2}\sigma_{2t}+\mathcal{H}\frac{d}{d\tau}\sigma_{2t}-
\frac{3}{2}\Omega_m(\tau)\mathcal{H}^2\sigma_{2t} \nonumber \\
&=&-\frac{3}{2}\Omega_m(\tau)\mathcal{H}^2
\delta_{1s}\delta_{1t}
+\bigg(\frac{\partial^i\partial^j}{\nabla^2}\delta_{1s}\bigg)
t_{ij}(\tau),
\end{eqnarray}
where $\sigma=\mr{s}^i_{\ ,i}$. Note that at linear order $\mr{M}_{1t}$
should not be included when expanding Eq. (\ref{eq5}), as it is induced by the 
long-wavelength tidal field $t_{ij}$, while the coupling between $\mr{M}_{1s}$ 
and $\mr{M}_{1t}$ should be included when expanding to second order as they 
source the local gravitational potential $\Phi_s$.
The first term on the right-hand side of Eq. (\ref{eq18}) vanishes since 
$\delta_{1t}=0$.
This equation can be solved numerically to get 
\begin{eqnarray}
\sigma_{2t}(\bm{q},\tau)=D_{\sigma1}(\tau)
\bigg(\frac{\partial^i\partial^j}{\nabla^2}\delta_{1s}(\bm{q},\tau)
\bigg)t_{ij}^{(0)},
\end{eqnarray}
where
\begin{equation}
D_{\sigma1}(\tau)=
\int^\tau_0d\tau'
\frac{H(\tau)D(\tau')-H(\tau')D(\tau)}{\dot{H}(\tau')D(\tau')-
H(\tau')\dot{D}(\tau')}\frac{T(\tau')D(\tau')}{D(\tau)}.
\end{equation}

The difference between density contrasts with and without $t_{ij}$ at $\bm{x}$ 
is 
\begin{eqnarray}
\delta_{t}(\bm{x})=\delta(\bm{x})-\delta(\bm{x})|_{t_{ij}=0}.
\end{eqnarray}
Since the tidal field induces the displacement $\mb{s}_{t}$ 
in addition to $\mb{s}_{s}$, the same Lagrangian coordinate $\bm{q}$ corresponds
to different $\bm{x}$ in these two cases. In the presence of $t_{ij}$, 
$\bm{x}=\bm{q}+\mb{s}_{1s}+\mb{s}_{1t}$. Here $\delta_{1s}$ solved from
linear equations gives the density at 
$\bm{x}_s=\bm{q}+\mb{s}_{1s}=\bm{x}-\mb{s}_{1t}$.
Taking this into account and using Eq. (\ref{eq3}), one finally obtain
\begin{eqnarray}
\delta_t(\bm{x})&=&\delta_{1t}(\bm{x})-\sigma_{2t}(\bm{x})+
\delta_{1t}(x)\delta_{1s}(\bm{x}) \nonumber \\
&+&\mr{tr}(\mr{M}_{1s}\mr{M}_{1t})|_{\bm{x}}
-\mr{s}^i_{1t}\partial_i \delta_{1s}(\bm{x}).
\end{eqnarray}
Inserting the first- and second-order solutions and noting that the tidal field
is traceless, we get
\begin{eqnarray}
\label{eq22}
\delta_t(\bm{x},\tau)=t^{(0)}_{ij}\bigg[
\alpha(\tau)\frac{\partial^i\partial^j}{\nabla^2}+\beta(\tau)x^i\partial^j
\bigg]\delta_{1s}(\bm{x},\tau),
\end{eqnarray}
where $\alpha(\tau)=-D_{\sigma1}(\tau)+F(\tau)$ and $\beta(\tau)=F(\tau)$.
The long-wavelength tidal field $t_{ij}$ leads to anisotropic small-scale
density fluctuations.

We have used the linear density-displacement relation 
$\delta_{1s}=-\mr{s}_{1s,i}^{\ i}$ repeatedly to convert the linear displacement
field $\mb{s}_{1s}$ to the small-scale density field $\delta_{1s}$. 
However, the linear relation holds badly at low redshifts and small scales
\cite{2012:log}. 
While the logarithmic relation between the divergence
of the displacement field and the density field,
\begin{eqnarray}
\mr{s}_{1s,i}^{\ i} = -\mr{ln}(1+\delta_{1s})+\langle\mr{ln}(1+\delta_{1s})\rangle,
\end{eqnarray}
is significantly better at low redshifts and small scales \cite{2012:log}.
Below we use the logarithmic variable $\delta_g=\mr{ln}(1+\delta_{1s})$ in 
place of $\delta_{1s}$ in the reconstruction. This
logarithmic transform reduces non-Gaussianities in the density field, hence 
improves the reconstruction \cite{2012:pen}.

The tidal field $t_{ij}$ with wave number $k_L$ can be taken as constant in a 
small patch with scale $\ll1/k_L$. Note it is different to have a constant 
tidal field $t_{ij}$ and a constant gravitational field $\Phi_L$.
The former corresponds to the second spatial derivative of the latter.
The local correlation function in the free-falling frame is 
\begin{eqnarray}
\xi(\bm{r},\tau)=\langle\delta(\bm{0},\tau)\delta(\bm{r},\tau)\rangle,
\end{eqnarray}
where $\delta=\delta_{1s}+\delta_{t}$. Using Eq. (\ref{eq22}), we obtain
\begin{eqnarray}
\label{eq25}
\xi(\bm{r},\tau)&=&\xi_{1s}(r,\tau)\nonumber\\
&+&t_{ij}^{(0)}\bigg[2\alpha(\tau)\frac{\partial^i\partial^j}{\nabla^2}+
\beta(\tau)r^i\partial^j\bigg]\xi_{1s}(r,\tau),
\end{eqnarray}
where $\xi_{1s}(r,\tau)=\langle\delta_{1s}(\bm{0},\tau)
\delta_{1s}(\bm{r},\tau)\rangle$ is the isotropic linear matter correlation
function. 
The anisotropic distortion of the locally measured small-scale density 
correlation function is induced by the long-wavelength tidal field $t_{ij}$.
Transforming Eq. (\ref{eq25}) to Fourier space, we get the local small-scale
power spectrum,
\begin{equation}
\label{eq26}
P(\bm{k},\tau)|_{t_{ij}}=P_{1s}(k,\tau)+
\hat{k}^i\hat{k}^jt_{ij}^{(0)}P_{1s}(k,\tau)f(k,\tau),
\end{equation}
where $P_{1s}(k,\tau)$ is the isotropic linear power spectrum and
\begin{equation}
\label{eq28}
f(k,\tau)=2\alpha(\tau)-\beta(\tau)\frac{{d\mr{ln}P_{1s}(k,\tau)}}{d\mr{ln}k}.
\end{equation}
Here, we abbreviate $k_S$ as $k$ and suppress the 
argument $k_L$ since $T(\tau)$ is scale independent.

Here, we only consider the leading-order coupling between the long-wavelength
tidal field and small-scale density fluctuations. However, in reality the 
density field is quite nonlinear and involves all higher-order interactions.
The estimators based on Eq. (\ref{eq26}) would be biased when the theoretical
description does not match the real situation. We address this problem
below.

\section{Tidal Reconstruction}
\label{sec:recon}

In this section, we present the algorithm for reconstructing the density field,
then we construct the tidal shear estimators.

\subsection{Reconstruction algorithm}
\label{sec:3b}
The tensor $\Phi_{L,ij}$ has six components. Such a symmetric $3\times3$ tensor 
can be decomposed into six independently observable components 
\cite{1973gw,2012:jeong}, $\Phi_{L,ij}=A_a\epsilon^a_{ij}$, where $A_a$ is
the expansion coefficient and $\epsilon^a_{ij}$ satisfies the orthogonal
relation $\epsilon^a_{ij}\epsilon^{bji}\propto\delta^{ab}$. We decompose
$\Phi_{L,ij}$ as
\begin{eqnarray}
\Phi_{L,ij}
&=&\kappa_\mr{3D}\delta_{ij}+t_{ij},
\end{eqnarray}
where $\kappa_\mr{3D}=\nabla^2\Phi_L/3$ and  
\begin{eqnarray}
\label{eq:tij}
t_{ij}=\left( \begin{array}{ccc}
\gamma_{1}-\gamma_{z} & \gamma_{2} & \gamma_{x}\\
\gamma_{2} & -\gamma_{1}-\gamma_{z} & \gamma_{y}\\
\gamma_{x} & \gamma_{y} & 2\gamma_z
\end{array} \right),
\label{eqn:tidalshear}
\end{eqnarray}
with $\gamma_{1}=(\Phi_{L,11}-\Phi_{L,22})/2$,
$\gamma_{2}=\Phi_{L,12}$, $\gamma_{x}=\Phi_{L,13}$, 
$\gamma_{y}=\Phi_{L,23}$, and 
$\gamma_{z}=(2\Phi_{L,33}-\Phi_{L,11}-\Phi_{L,22})/6$. 
The tensor is decomposed in a way such that when reduced 
to the 2D case notations reduce to those used in gravitational lensing, 
so the third axis plays a different role than the other two axes.
For intuitive understandings of these abstract symbols, see the figures in 
Ref. \cite{2012:jeong}. All these six different components will induce
different observable effects in the local quadratic statistics, including the
isotropic modulations ($\kappa_\mr{3D}$) and anisotropic parts ($\gamma_1$, 
$\gamma_2$, etc). The large-scale gravitational potential $\Phi_L$ 
 is a single number, so it is sixfold overdetermined. 
Using the isotropic modulation to reconstruct the large-scale density field 
(supersample signal) has been studied in Ref. \cite{2014li}. 
The local halo bias discussed in Refs. 
\cite{2016PhRvD..93f3507L,2016JCAP...02..018L,2015arXiv151101465B} recently 
arises from the isotropic modulation on the distribution of halos.
Here, we focus on the ``change of shape", i.e., the traceless tidal field 
$t_{ij}=\Phi_{L,ij}-\nabla^2\Phi_{L}\delta_{ij}/3$.

The induced local anisotropy pattern by the long-wavelength tidal field is 
proportional to $t_{ij}$. Using orthogonal components introduced above, 
Eq.(\ref{eq26}) can be written as
\begin{equation}
\frac{P(\bm{k},\tau)|_{t_{ij}}}{P_{1s}(k,\tau)}-1=f(k,\tau)
[(\hat{k}_1^2-\hat{k}_2^2)\gamma_1^{(0)}+
2\hat{k}_1\hat{k}_2\gamma_2^{(0)}]+ \cdots
\end{equation}
Here, we only write out the $\gamma_1$ and $\gamma_2$ terms explicitly.
The distortion along the line of sight is affected by peculiar velocities in 
spectroscopic surveys, so here we shall use $\gamma_1$ and $\gamma_2$, which 
only involve derivatives in the tangential plane.
In the language of weak lensing, $\gamma_1$ and $\gamma_2$ describe
quadrupolar distortions in the tangential plane.
Peculiar velocities cause particles moving along $x_\parallel$ axis, but we 
expect the changes in $\gamma_1$ and $\gamma_2$ due to peculiar velocities to
be a second-order effect.

We can obtain the 2D convergence field 
$\kappa_\mr{2D}=(\Phi_{L,11}-\Phi_{L,22})/2$ from $\gamma_1$ and 
$\gamma_2$ as in gravitational lensing \cite{1993kaiser}, 
$\kappa_{\mr{2D},11}+\kappa_{\mr{2D},22}=
\gamma_{1,11}-\gamma_{1,22}+2\gamma_{2,12}$.
In Fourier space, this can be written as
\begin{eqnarray}
\kappa_\mr{2D}(\bm{k})=\frac{1}{k_1^2+k_2^2}[(k_1^2-k_2^2)\gamma_1(\bm{k})
+2k_1k_2\gamma_2(\bm{k})].
\end{eqnarray}
The 2D convergence can be converted into the 3D convergence field $\kappa_\mr{3D}=\nabla^2\Phi_L/3$ as
\begin{eqnarray}
\label{eq:kappa}
\kappa_\mr{3D}(\bm{k})&=&\frac{2k^2}{3(k_1^2+k_2^2)}\kappa_\mr{2D}(\bm{k})\\
&=&\frac{2k^2}{3(k_1^2+k_2^2)^2}
[({k}_{1}^2-{k}_{2}^2)\gamma_1(\bm{k})
+2{k}_{1}{k}_{2}\gamma_2(\bm{k})].\nonumber
\end{eqnarray}
Now we get an estimate of the large-scale density field by measuring $\gamma_1$
and $\gamma_2$. The large-scale density field $\delta_L$ only differs from
$\kappa_\mr{3D}$ by a constant factor which we address in a moment.

This reconstruction is inherently 3D instead of 2D. We only use $\gamma_1$ and 
$\gamma_2$, so one might have thought that the cosmic 
tidal reconstruction takes place in a single 2D slice.
However, although $\gamma_1$ and $\gamma_2$ only involve derivatives in the 
tangential plane, the changes of $\gamma_1$ and $\gamma_2$ along $x_\parallel$ 
axis encode the change of $\delta_L$ along $x_\parallel$ axis.
Since we only use $\gamma_1$ and $\gamma_2$ instead of all components in the 
tidal field $t_{ij}$ for reconstruction, the variance of the reconstructed 
mode $\kappa_\mr{3D}(\bm{k})$ is anisotropic in $\bm{k}$. 
The variance of $\kappa_\mr{3D}(\bm{k})$ is 
\begin{eqnarray}
&&\langle\kappa_\mr{3D}(\bm{k})\kappa_\mr{3D}(\bm{k}')\rangle=
\big[({k}_{1}^2-{k}_{2}^2)({k_1'}^2-{k_2'}^2)
\langle\gamma_1(\bm{k})\gamma_1(\bm{k}')\rangle\nonumber\\
&&+(2{k}_{1}{k}_{2})(2{k_1'}{k_2'})
\langle\gamma_2(\bm{k})\gamma_2(\bm{k}')\rangle\big]
\times\frac{2k^2}{3(k_\perp^2)^2}\frac{2{k'}^2}{3({k_\perp'}^2)^2},
\end{eqnarray}
where $k_\perp=k_1^2+k_2^2$. 
The power spectra of $\gamma_1$ and $\gamma_2$ are 
scale independent on large scales \cite{1999matias}, so
\begin{eqnarray}
&&\langle\kappa_\mr{3D}(\bm{k})\kappa_\mr{3D}(\bm{k}')\rangle \propto 
\big[({k}_{1}^2-{k}_{2}^2)({k_1'}^2-{k_2'}^2)\nonumber\\
&&+(2{k}_{1}{k}_{2})(2{k_1'}{k_2'})\big]
\times\frac{2k^2}{3(k_\perp^2)^2}\frac{2{k'}^2}{3({k_\perp'}^2)^2}
\delta^D(\bm{k}+\bm{k}'),
\end{eqnarray}
where $\delta^D(\bm{k})$ is the Dirac delta function.  The
noise in the reconstructed mode $\kappa_\mr{3D}(\bm{k})$ is 
$\sigma^2_{\kappa_\mr{3D}}\propto (k^2/k_\perp^2)^2$, which is anisotropic in 
$k_\perp$ and $k_\parallel$.
If we include the components that include $i=3$, the noise would be isotropic, 
but redshift space distortions will influence the reconstruction. 
We shall investigate this in the future. 

We see that when $k_\perp\to0$ the noise is infinite for the corresponding 
$\kappa_\mr{3D}(\bm{k})$. In this case, the estimator in Eq.(\ref{eq:kappa}) 
diverges. As these modes contain nothing but noise, we set these modes to zero 
in the reconstruction.
Since the noises are different for modes with with different $k_\perp$ and 
$k_\parallel$, we need to filter the reconstructed density field 
$\kappa_\mr{3D}$ to obtain the  clean density field $\kappa$. 

In general, the reconstructed noisy 3D convergence field $\kappa_\mr{3D}$  
can be written as
\begin{eqnarray}
\label{eq:kap3d}
\kappa_\mr{3D}(k_\perp, k_\parallel) = b(k_\perp,k_\parallel)
\delta_L(k_\perp, k_\parallel) + n(k_\perp, k_\parallel),
\end{eqnarray}
where $b(k_\perp, k_\parallel)$ is the bias factor and $n(k_\perp,k_\parallel)$
is the noise in reconstruction. The bias factor and the noise can be determined
from the cross correlation of $\kappa_\mr{3D}$ and 
$\delta$ and the correlation function of $\kappa_\mr{3D}$,
\begin{equation}
\langle\kappa_\mr{3D}\delta\rangle=b\langle\delta\delta\rangle, \qquad
\langle\kappa_\mr{3D}\kappa_\mr{3D}\rangle=b^2\langle\delta\delta\rangle+
\langle nn\rangle.
\end{equation}
We then get 
\begin{eqnarray}
\label{eq:bias}
b(k_\perp, k_\parallel) = \frac{P_{\kappa_\mr{3D}\delta}(k_\perp,k_\parallel)}{
P_{\delta}(k_\perp,k_\parallel)},
\end{eqnarray}
and
\begin{eqnarray}
\label{eq:psnoise}
P_n(k_\perp,k_\parallel)=
P_{\kappa_\mr{3D}}(k_\perp,k_\parallel)
-b^2(k_\perp,k_\parallel)P_{\delta}(k_\perp,k_\parallel).
\end{eqnarray}
We correct the bias factor and apply the Wiener filter to obtain the 
reconstructed clean field, 
\begin{eqnarray}
\label{eq:kap}
{\kappa}(k_\perp,k_\parallel)=\frac{\kappa_\mr{3D}(k_\perp,k_\parallel)}{b(k_\perp,k_\parallel)} 
W(k_\perp,k_\parallel),
\end{eqnarray}
where
\begin{eqnarray}
W(k_\perp,k_\parallel)=\frac{P_{\delta}(k_\perp,k_\parallel)}
{P_{\delta}(k_\perp,k_\parallel)+P_{n}(k_\perp,k_\parallel)/b^2
(k_\perp,k_\parallel)}.
\end{eqnarray}

The bias factor $b(\bm{k})$ accounts two biasing effects. One is because that
the leading-order perturbation theory does not describe the full nonlinear 
process [$b_\mr{NL}(\bm{k})$]. Another is because we use the estimators 
derived in the
long-wavelength limit [$b_\gamma(\bm{k})$], which we discuss in a moment.

\subsection{Tidal shear estimators}
We first construct the unbiased minimum variance tidal shear estimators
in the long-wavelength limit, where $\gamma_1$ and 
$\gamma_2$ are constant in space, then generalize the estimators to the 
spatial varying case. We also discuss what happens when the long-wavelength 
limit does not hold.

In the long-wavelength limit and under the Gaussian assumption, 
quadratic estimators can be constructed either by using the maximum likelihood method\cite{2008:lu}
or the inverse variance weighting \cite{2010lu,2012bucher},
\begin{equation}
\hat{\gamma}_1=\frac{1}{Q_{\gamma_1}}\int\frac{d^3k}{(2\pi)^3}
\frac{|{\delta}_g(\bm{k})|^2}{L^3}
\frac{P(k)}{P_{\mr{tot}}^2(k)}f(k)(\hat{k}_1^2-\hat{k}_2^2),
\end{equation}
\begin{equation}
\hat{\gamma}_2=\frac{1}{Q_{\gamma_2}}\int\frac{d^3k}{(2\pi)^3}
\frac{|{\delta}_g(\bm{k})|^2}{L^3}
\frac{P(k)}{P_{\mr{tot}}^2(k)}f(k)(2\hat{k}_1\hat{k}_2),
\end{equation}
where $P_\mr{tot}(k)=P(k)+P_\mr{N}(k)$ is the observed power spectrum,
including both the signal and noise, and
\begin{eqnarray}
Q_{\gamma_1}=\int\frac{d^3k}{(2\pi)^3}
\frac{P^2(k)}{P_{\mr{tot}}^2(k)}f^2(k)(\hat{k}_1^2-\hat{k}_2^2)^2,
\end{eqnarray}
\begin{eqnarray}
Q_{\gamma_2}=\int\frac{d^3k}{(2\pi)^3}
\frac{P^2(k)}{P_{\mr{tot}}^2(k)}
f^2(k)
(2\hat{k}_1\hat{k}_2)^2.
\end{eqnarray}
The factor $L^3$ appears because we consider the density field defined in a 
finite volume of side length $L$.
After integrating over angles in Fourier space, we have
\begin{equation}
Q_{\gamma_1}=Q_{\gamma_2}=Q=\int\frac{2k^2dk}{15\pi^2}\frac{P^2(k)}{P_{\mr{tot}}^2(k)}
f^2(k)
\label{eq:Q}
\end{equation}
Using Parseval's theorem, we can rewrite the  above equations in real space,
\begin{eqnarray}
\label{eq36}
\hat{\gamma}_1=\frac{1}{L^3}\int\d^3x
\big[{\delta}^{w_1}_g(\bm{x}){\delta}^{w_1}_g(\bm{x})-
{\delta}^{w_2}_g(\bm{x}){\delta}^{w_2}_g(\bm{x})\big],
\end{eqnarray}
and
\begin{eqnarray}
\label{eq37}
\hat{\gamma}_2=\frac{1}{L^3}\int\d^3x
\big[2{\delta}^{w_1}_g(\bm{x}){\delta}^{w_2}_g(\bm{x})\big],
\end{eqnarray}
where ${\delta}^{w_1}_g(\bm{x})$ and ${\delta}^{w_2}_g(\bm{x})$ 
are two filtered density fields. In Fourier space, they are given by
\begin{eqnarray}
{\delta}_g^{w_i}(\bm{k})=\delta_g(\bm{k})w_i(\bm{k}),
\end{eqnarray}
where 
\begin{eqnarray}
\label{eq:filter}
w_i(\bm{k})=\bigg[\frac{P(k)f(k)}{QP^2_\mr{tot}(k)}\bigg]^{1/2}
i\hat{k}_i.
\end{eqnarray}

In deriving Eqs. (\ref{eq36}) and (\ref{eq37}), we have assumed that the
tidal shear field is constant in space. 
From the expressions for $\hat{\gamma}_1$ and $\hat{\gamma}_2$, we can 
see the terms in square brackets give estimates for $\gamma_1$ and 
$\gamma_2$ at the location $\bm{x}$, while the integral $\int d^3x/L^3$ 
averages the local value over the whole space $L^3$.
If the fluctuation of the tidal field $t_{ij}$ is slow compared to the filter, 
we can use the localized estimation in the square brackets as estimates
for $\gamma_1,\gamma_2$ \cite{2008:lu,2010lu,2012bucher}. 
So the estimators
\begin{eqnarray}
\label{eq40}
\hat{\gamma}_1(\bm{x})&=&
\big[{\delta}^{w_1}_g(\bm{x}){\delta}^{w_1}_g(\bm{x})-
{\delta}^{w_2}_g(\bm{x}){\delta}^{w_2}_g(\bm{x})\big],\nonumber\\
\hat{\gamma}_2(\bm{x})&=&
\big[2{\delta}^{w_1}_g(\bm{x}){\delta}^{w_2}_g(\bm{x})\big],
\end{eqnarray}
provide the unbiased minimum variance estimates of the spatial varying tidal
shear fields in the long-wavelength limit.

The estimators derived in Eq.(\ref{eq40}) assume the long-wavelength tidal shear
fields vary slowly compared to the small-scale density field.
If the long-wavelength limit is not satisfied, the tidal shear reconstruction 
would be biased by a multiplicative bias factor $b_\gamma(\bm{k})$, which 
approaches to unity in the limit $k\to0$ and decreases when $k$ increases 
\cite{2008:lu,2012bucher}. 
The bias factor can be derived in the case of Gaussian density perturbations 
\cite{2008:lu}. It has also been studied numerically in the cosmic microwave background lensing 
\cite{2012bucher}.
In the cosmic tidal reconstruction, we mainly use the nonlinear structures 
around the scale $1.25\ \mr{Mpc}/h$ to reconstruct the long-wavelength tidal 
field \cite{2012:pen}, so for the reconstructed large-scale density 
perturbations with wave number $k_L\lesssim0.1\ h$/Mpc we can ignore this 
multiplicative bias.


\section{Simulation}
\label{sec:sim}

In this section, we explore the cosmic tidal reconstruction process in all 
details using the dark matter density fields from $N$-body simulations. 
Then, we discuss the dependence on smoothing scale and the necessity of 
Gaussianization. At last, we study the anisotropic noise in reconstruction.

We run $N$-body  simulations using the CUBE$\mr{P}^3$M code \cite{2013:code} 
with $1024^3$ dark matter particles in a box of side length $L=1.2\ \mr{Gpc}/h$.
We have adopted the following set of cosmological parameter values:
$\Omega_b=0.049$, $\Omega_c=0.259$, $h=0.678$,
$A_s=2.139\times10^{-9}$ and $n_s=0.968$.  Six runs with independent 
initial conditions were performed to provide better statistics.
In the following calculations, we use outputs at $z=0$.

\subsection{An example of cosmic tidal reconstruction}
In this subsection, we present a complete tidal reconstruction process. We  
follow the simple, slightly suboptimal scenario in Ref. \cite{2012:pen}. 
We first smooth the 3D density field using a Gaussian window function,
\begin{eqnarray}
\bar{\delta}(\bm{x})=\int d^3x'S(\bm{x}-\bm{x'})\delta(\bm{x}'),
\end{eqnarray}
where $S(\bm{r})=e^{-r^2/2R^2}$. The smoothing scale is $R=1.25 \mr{Mpc}/h$ 
as in Ref. \cite{2012:pen}. In the following subsection, we demonstrate the reconstruction result is 
not sensitive to the smoothing scale unless we smooth on linear scales, i.e.,
$R\gtrsim 5\ \mr{Mpc}/h$. Then, we take a logarithmic transform
\begin{eqnarray}
\delta_g(\bm{x})=\mr{ln}[1+\bar{\delta}(\bm{x})]
\end{eqnarray}
to Gaussianize the density field as motivated by Ref. \cite{2012:log}.
The additive constant $\langle\mr{ln}(1+\bar{\delta}(\bm{x}))\rangle$ will 
not affect our reconstruction results, as we only use the derivatives of this
field.

Next, we convolve the logarithmic density field $\delta_g(\bm{x})$ with the 
filter $w_i(\bm{x})$ in Eq. (\ref{eq:filter})
and then obtain the three-dimensional
tidal shear fields $\gamma_1(\bm{x})$ and $\gamma_2(\bm{x})$ using
Eq. (\ref{eq40}). 
The two tidal shear fields $\gamma_1(\bm{x})$ and $\gamma_2(\bm{x})$ directly
describe the quadrupolar distortions of the density field in the tangential
plane, while the variations of the density field along the $x_\parallel$ axis 
are inferred from the variations of the transverse tidal shear fields in the 
$x_\parallel$ direction. 

By combining $\gamma_1(\bm{k})$ and $\gamma_2(\bm{k})$
using Eq. (\ref{eq:kappa}), we obtain the 3D convergence field 
$\kappa_\mr{3D}(\bm{k})$. We estimate $\langle\kappa_\mr{3D}\delta\rangle$ and 
$\langle\kappa_\mr{3D}\kappa_\mr{3D}\rangle$ using the power spectra from these 
six simulations. Then, from Eqs. (\ref{eq:bias}) and (\ref{eq:psnoise}) we 
get the bias factor $b(k_\perp, k_\parallel)$ and the noise power spectrum 
$P_n(k_\perp,k_\parallel)$. 
The clean field ${\kappa}(\bm{k})$ is given by Eq. (\ref{eq:kap}).

\begin{figure*}[tbp]
\begin{center}
\includegraphics[width=0.95\textwidth]{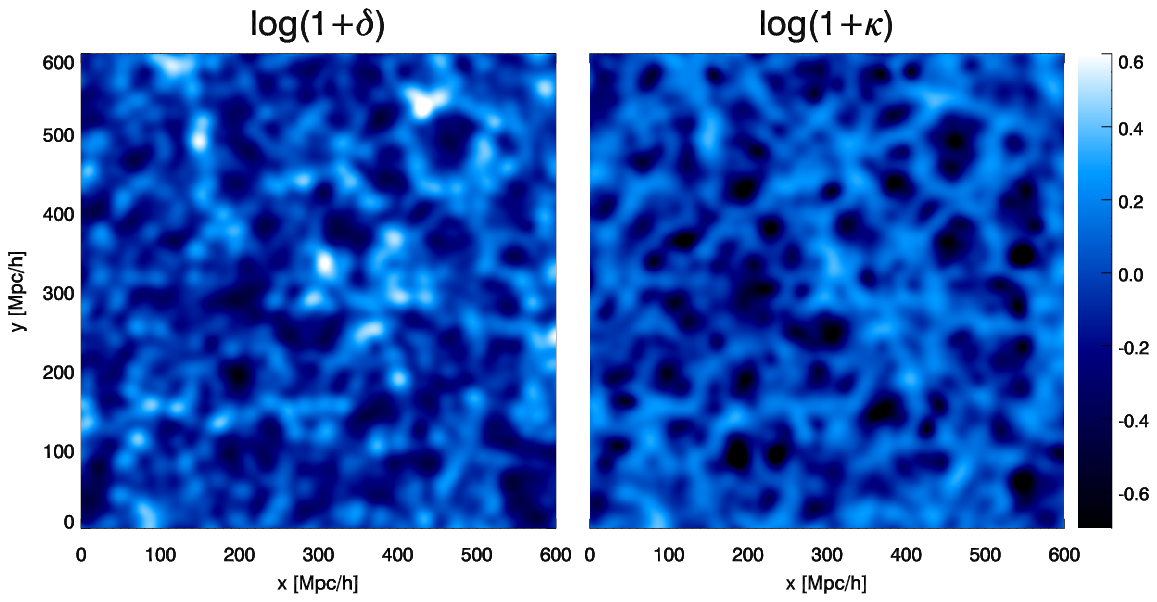}
\end{center}
\caption{\label{fig:denxy} The left panel shows a slice of the original density
field $\delta(\bm{x})$ smoothed on 8 Mpc/$h$ in $x-y$ plane. 
The right panel shows the corresponding slice of the reconstructed density 
field $\kappa(\bm{x})$, also smoothed on 8 Mpc/$h$. }
 \end{figure*}

\begin{figure*}[tbp]
\begin{center}
\includegraphics[width=0.95\textwidth]{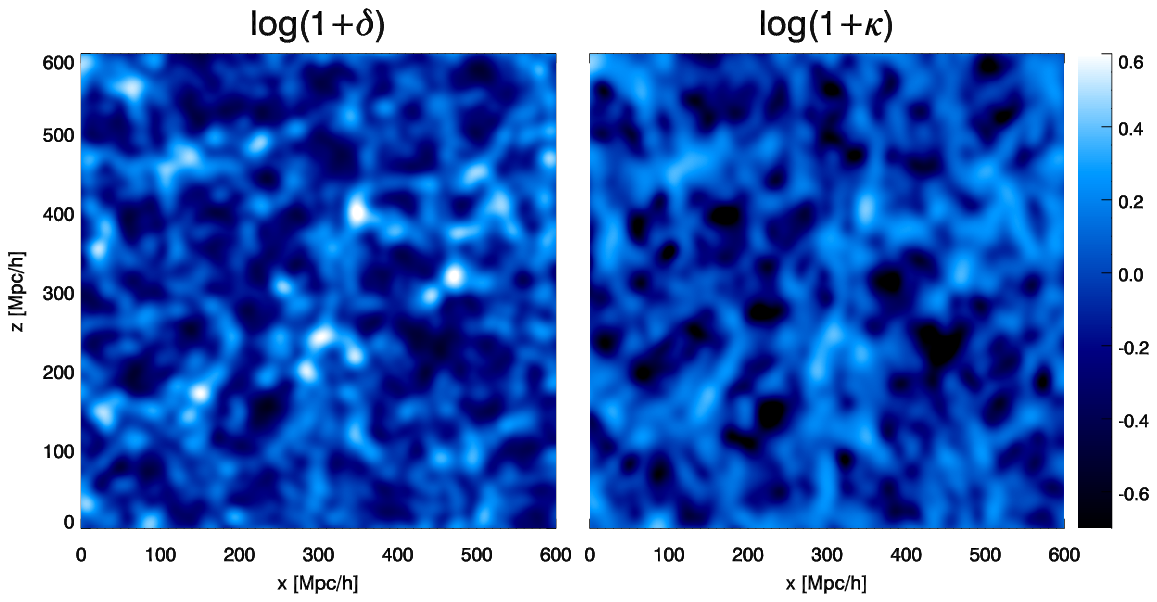}
\end{center}
\caption{\label{fig:denxz} 
The left panel shows a slice of the original density field $\delta(\bm{x})$ smoothed on 8 Mpc/$h$ in $x-z$ plane. 
The right panel shows the corresponding slice of the reconstructed density 
field $\kappa(\bm{x})$, also smoothed on 8 Mpc/$h$. }
\end{figure*}

In Fig. \ref{fig:denxy}, we show a $1.17\ \mr{Mpc/}h$ slice of the original
density field in the $x-y$ plane in the left panel and a slice of the 
reconstructed density field in the right panel, both smoothed on $8\ \mr{Mpc}/h$
to reduce the small-scale noise in order to make visual comparisons.  
In Fig. \ref{fig:denxz}  we show the similar slices in the $x-z$ plane.   
We find in both cases the reconstructed density fields resemble the original
ones. However, the strong peaks in the original density 
field are less prominent in the reconstructed field.
The peaks in the original density field correspond to strongly nonlinear 
structures, like some very massive halos. 
In deriving Eq.(\ref{eq26}), we assume that the small-scale density fluctuations
undergo linear evolutions and the local anisotropies are due to the long-wavelength
tidal field. 
However, the self-gravitational interaction is very strong around these 
nonlinear 
structures, so they change little under the weak gravitational interaction 
with the long-wavelength tidal field.
Equation (\ref{eq26}) does not hold in these regimes.
The reconstruction exploits the nonlinear coupling between the long-wavelength
tidal field and the small-scale density fluctuations but limited by the strong
nonlinear gravitational clustering on small scales during structure formation,
which leads to non-Gaussianity.
Indeed, the original density field needs to be smoothed and Gaussianized to 
reduce the weighting of such contributions in order to obtain  
good reconstruction results.

\begin{figure}[tbp]
\begin{center}
\includegraphics[width=0.48\textwidth]{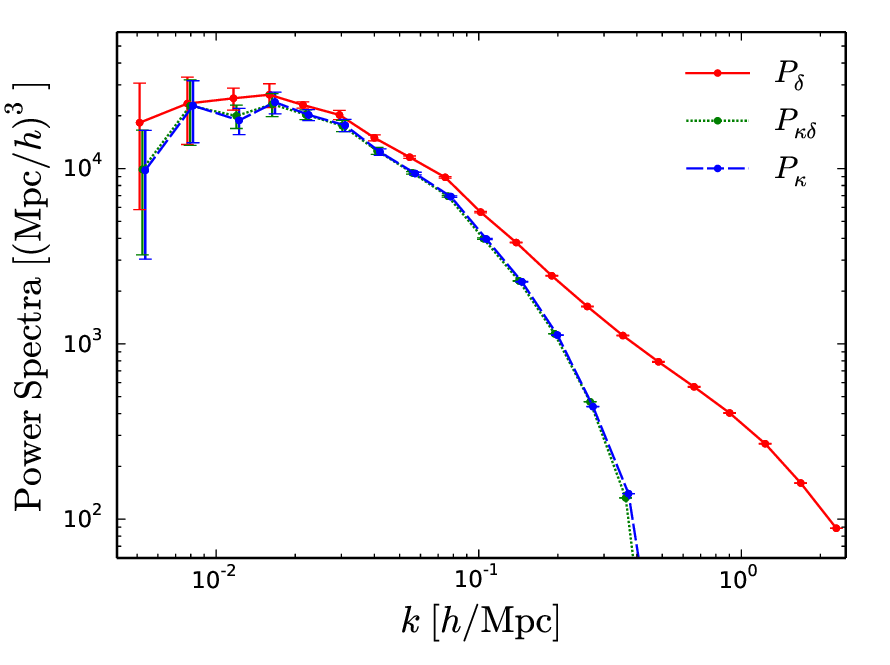}
\end{center}
\vspace{-0.7cm}
\caption{\label{fig:power} The auto and cross power spectra of the original and reconstructed density fields. 
}
\end{figure}

In Fig. \ref{fig:power}, we show the auto power spectra of the original field
$\delta$, the reconstructed field $\kappa$ and the cross power spectrum between 
$\kappa$ and $\delta$. 
Data points in the same $k$-bin are shifted slightly for clarity of display. 
The large errors at $k\sim0.005\ h/\Mpc$ are due to finite modes on the largest scales.
We see a good cross correlation between the original and 
reconstructed density fields over a wide range of wave numbers, 
$0.008\ h/\Mpc \lesssim k \lesssim 0.1\ h/\Mpc$.
The general shape of $P_{\kappa}$ is similar to that of $P_\delta$. 
The power spectrum for $\kappa$ and the cross power spectrum between 
$\kappa$ and $\delta$ are nearly the same in this range because 
after we apply the Wiener filter to the reconstructed noisy field 
$\kappa_\mr{3D}$, the part correlated with $\delta$ remains
in the clean field $\kappa$.
At $k>0.1\Mpc/h$,  the auto power spectrum of $\kappa$ and the cross power 
spectrum drop more rapidly than the original density field.

In Fig. \ref{fig:coeff}, we plot the cross-correlation coefficient 
between the original density field and reconstructed density field, 
\begin{equation}
r_{\kappa\delta} = {P_{\kappa\delta}}/{\sqrt{P_{\delta}P_{\kappa}}}.  
\end{equation}
The error bars are estimated by the bootstrap resampling method. 
The cross-correlation coefficient $r_{\kappa\delta}$ is larger than 0.9 for 
$0.008\ h/\Mpc \lesssim k \lesssim 0.1\ h/\Mpc$.
The cross-correlation coefficient drops rapidly at $k>0.1\ \Mpc/h$ because  
the reconstruction ceases to work on small scales. 

\subsection{Dependence on smoothing scales and Gaussianization methods}
\begin{figure}[tbp]
\begin{center}
\includegraphics[width=0.48\textwidth]{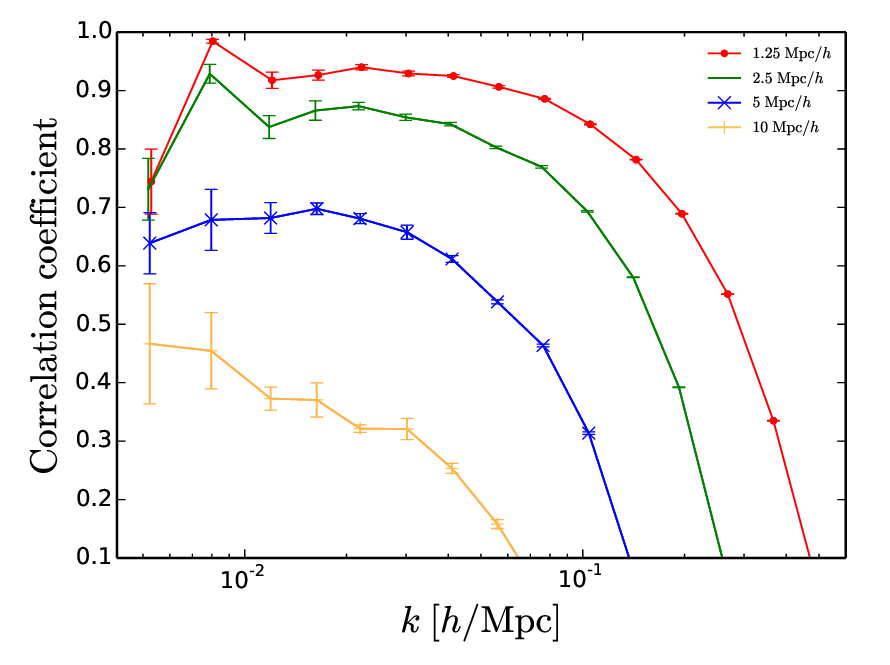}
\end{center}
\vspace{-0.7cm}
\caption{\label{fig:coeff} The cross-correlation coefficients for four different
smoothing scales, $R=1.25$, 2.5, 5, and 10 $\mr{Mpc}/h$. The cross-correlation 
coefficients decrease when the smoothing scales are increased.}
\end{figure}

To quantify the dependence on the smoothing scale, we perform reconstruction 
with three more smoothing scales, $R=2.5$, 5, and $10\ \mr{Mpc}/h$. 
Figure \ref{fig:coeff} shows the cross-correlation coefficients.
The cross-correlation coefficients decrease when the smoothing scales are 
increased. We are using the small-scale structures to reconstruct the 
large-scale density field. When we increase the smoothing scale, we are losing 
the small-scale structures that can be used to provide cosmological information
about the large-scale density field. However, the reconstruction results do 
not degrade significantly until the smoothing scale goes beyond $5\ \mr{Mpc}/h$,
roughly the scale on which the structure growth transits from linear to 
nonlinear. The reconstruction is not sensitive to the smoothing scale unless 
we smooth on the linear scales.

\begin{figure}[tbp]
\begin{center}
\includegraphics[width=0.48\textwidth]{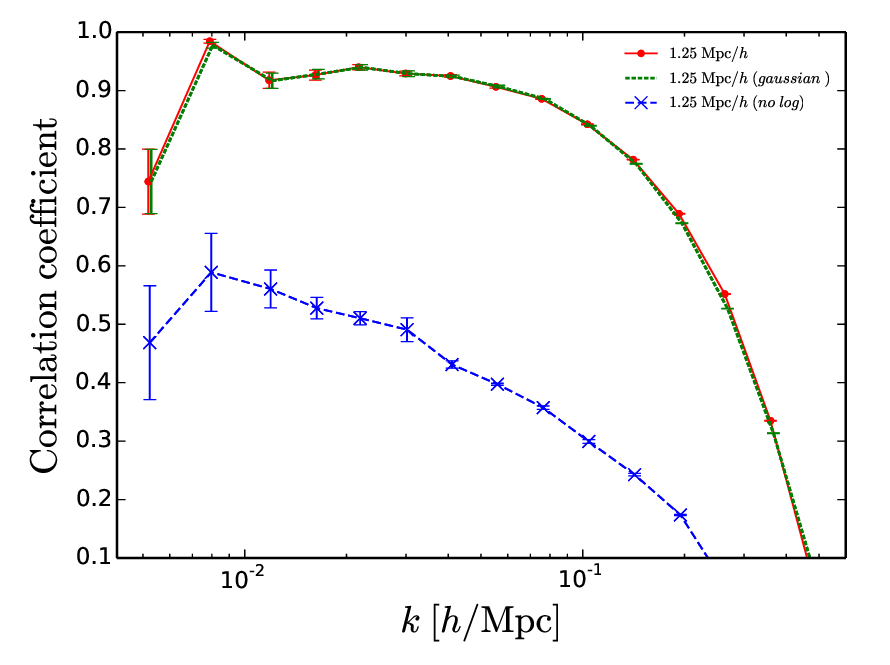}
\end{center}
\vspace{-0.7cm}
\caption{\label{fig:coeff2} The cross-correlation coefficients for different
Gaussianization methods. The solid line shows the reconstruction with the 
logarithmic transform. The dotted line shows the reconstruction using the normal
density field. The dashed line shows the reconstruction without the logarithmic
transform. The cross-correlation coefficients for the reconstruction with the
logarithmic transform and the normal density field are almost the same.}
\end{figure}

The logarithmic transform plays an important role in the cosmic tidal 
reconstruction. In Fig. \ref{fig:coeff2}, we also plot the result for a 
reconstruction without the logarithmic transform. 
The cross-correlation of the reconstructed field with the original 
field is much weaker than the one with the logarithmic transform and quickly 
drops to nearly zero when the wave number increases. 
Then, we adopt a different Gaussianization method by ranking the original density
fluctuations into a Gaussian distribution \cite{2009ApJ...698L..90N} and 
perform tidal reconstruction using this normal density field.
The cross-correlation coefficient is almost the same as that using the 
logarithmic transform.
Apparently, the non-Gaussianity limits the shear reconstruction. 
The tidal shear estimators derived under the Gaussian assumption are not 
necessarily optimal for the non-Gaussian density field. 
The logarithmic transform reduces the non-Gaussianity of the density field 
so that a better result is achieved. It also 
fixes the displacement-density relation significantly as shown in 
Ref. \cite{2012:log}.

\subsection{The anisotropic noise}
\begin{figure}[tbp]
\begin{center}
\includegraphics[width=0.48\textwidth]{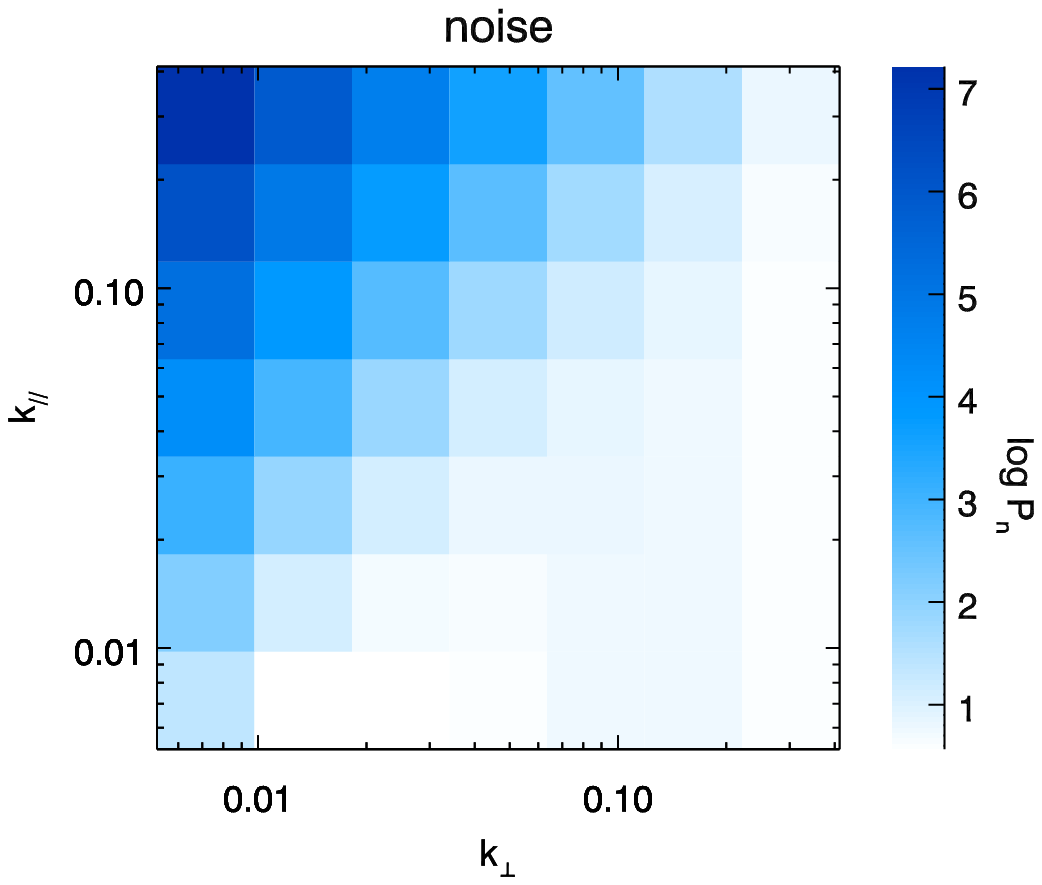}
\end{center}
\vspace{-0.7cm}
\caption{\label{fig:noise} The anisotropic noise power spectrum 
$P_n(k_\parallel, k_\perp)$. The noises for modes with large $k_\parallel$ and
small $k_\perp$ are orders of magnitude larger than other modes.
}
\end{figure}

\begin{figure}[tbp]
\begin{center}
\includegraphics[width=0.48\textwidth]{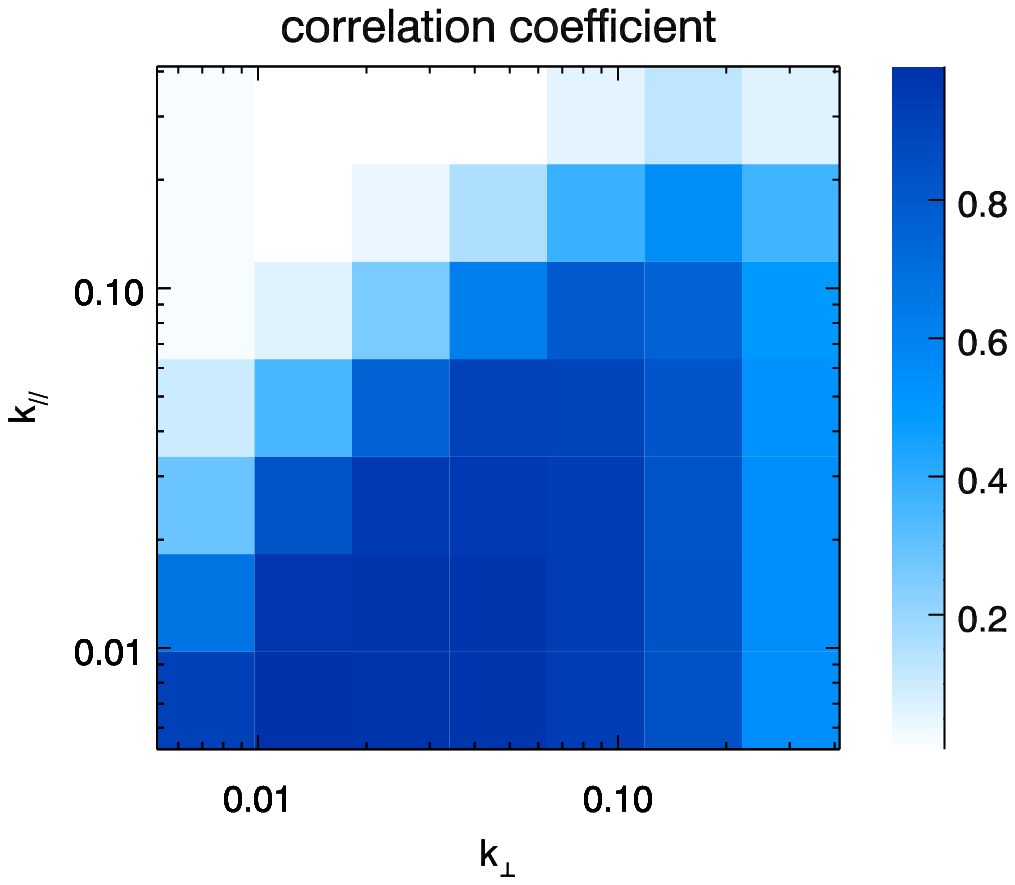}
\end{center}
\vspace{-0.7cm}
\caption{\label{fig:ratio} The cross-correlation coefficient 
$r_{\kappa\delta}(k_\parallel,k_\perp)$.  
The cross-correlation is better at the lower right corner, which can be close 
to 1 for modes with small $k_\parallel$, but drops quickly when $k_\parallel$ 
increases; the modes with large $k_\parallel$ and small $k_\perp$ are poorly 
reconstructed.
}
 \end{figure}

The noise of $\kappa_\mr{3D}(\bm{k})$ is anisotropic as we discussed earlier. 
We show the anisotropic noise power spectrum in Fig. \ref{fig:noise}. 
The noises for modes with large $k_\parallel$ and small $k_\perp$ are orders 
of magnitude larger than other modes. This result confirms the estimate of 
noise derived from theory, $\sigma^2_{\kappa_\mr{3D}}\propto (k^2/k_\perp^2)^2$,
which diverges for small $k_\perp$. In Fig. \ref{fig:ratio}, we show the 
anisotropic cross-correlation coefficient 
$r_{\kappa\delta}(k_\parallel,k_\perp)=P_{\kappa\delta}(k_\parallel,k_\perp)/\sqrt{P_{\delta}(k_\parallel,k_\perp)P_{\kappa}(k_\parallel,k_\perp)}$.
The correlation between the original and reconstructed fields is better 
at the lower right corner, which can be close to 1 for modes with small 
$k_\parallel$, but drops quickly when $k_\parallel$ increases; the modes with 
large $k_\parallel$ and small $k_\perp$ are poorly reconstructed.

\begin{figure}[tbp]
\begin{center}
\includegraphics[width=0.48\textwidth]{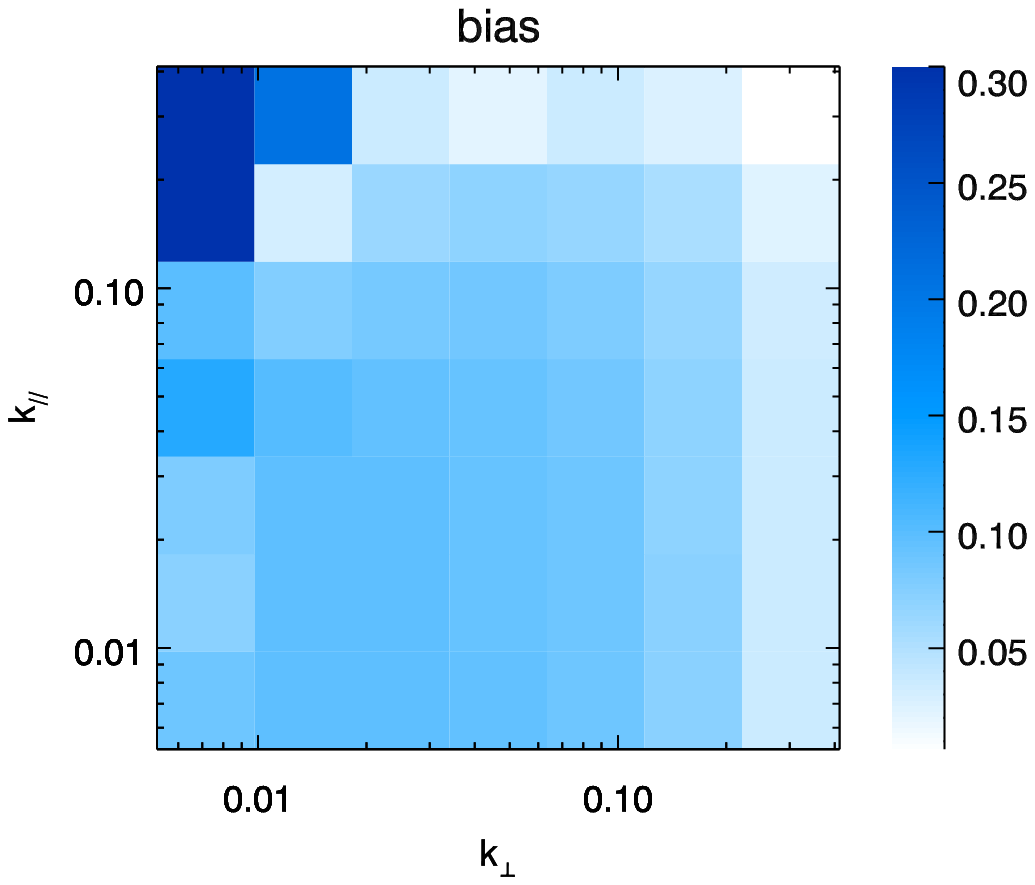}
\end{center}
\vspace{-0.7cm}
\caption{\label{fig:bias} The bias factor $b(k_\parallel,k_\perp)$. The bias 
factor is almost scale independent in this plane. 
$b(k_\parallel,k_\perp)$ saturates at the upper left corner. Since these
modes are noisy, the extremely large values of the bias at that corner are 
not reliable.}
\end{figure}

In Fig. \ref{fig:bias}, we show the anisotropic bias factor.  
The bias factor is nearly constant except for modes with very large 
$k_\parallel$ and very small $k_\perp$. Here,
$b(k_\parallel,k_\perp)$ saturates at the upper left corner. Since these
modes are noisy, the extremely large values of the bias at that corner are 
not reliable.
This constancy of the bias factor shows that our reconstructed density 
field is a faithful representation of the original density field.
It is also encouraging for applying the reconstructed density field to do 
cross correlations with other tracers for beating down cosmic variances 
\cite{2009PhRvL.102b1302S,2009mcdonald}. 

The anisotropy in reconstruction is due to the fact we only use tidal shear 
fields in the tangential plane. The changes of the long-wavelength density 
field $\delta_L$ along the line of sight are inferred indirectly from the 
variations of tidal shear fields $\gamma_1(\bm{x})$ and $\gamma_2(\bm{x})$ 
along $x_\parallel$ axis, so we cannot capture the rapid changes of the
density field along $x_\parallel$ axis. 
By including tidal shear fields containing derivatives with 
respect to $x_\parallel$ axis, the reconstruction will be improved. 
However, redshift space distortions will inevitably affect the reconstruction 
result. We leave the detailed study of this to a future paper.

\section{Discussions}
\label{sec:disc}

In this paper, we show how the long-wavelength tidal field $t_{ij}$ can be 
reconstructed from the local small-scale power spectrum. 
To reconstruct $t_{ij}$ from the anisotropic small-scale
density perturbations, we need to learn the interaction process clearly.
In this paper, we follow the treatments in Ref. \cite{2014:tidal}, only considering 
the coupling between $\mb{s}_{1s}$ and $\mb{s}_{1t}$, which is the leading-order
effect from $t_{ij}$. However, unlike weak lensing, where the interaction is
weak so that the linearized calculation is good enough for most cases, the 
tidal interaction is inherently a nonlinear process. 
Higher-order terms of form $(\mb{s}_{1s})^n\mb{s}_{1t}$ 
$(n>1)$ also play important roles in the tidal interaction, as the
nonlinearities on small scales are quite strong. 
The proportional coefficient in Eq. (\ref{eq28}) will change if we include 
terms like $(\mb{s}_{1s})^n\mb{s}_{1t}$ $(n>1)$. 
The tidal shear estimators we used in this paper are biased since we do 
not know the tidal interaction process accurately.  
However, even if we include some higher-order terms, the constructed estimators
might still be biased, as this still does not describe the full nonlinear 
process.
In this paper, we assume that the nonlinearities only change the proportional 
coefficient $f(k,\tau)$. The tidal distortion of the local power spectrum 
is still proportional to $\hat{k}^i\hat{k}^jt_{ij}^{(0)}$, and the proportional
coefficient does not vary significantly over different wave numbers.
Then, we can absorb the unknown coefficient into the bias factor introduced
in Eq. (\ref{eq:kap3d}). The bias factor can be solved from $N$-body
simulations. The proportional constant from the Poisson equation and the
normalization constant $Q$ in the tidal shear estimators can also be absorbed
in this bias factor.
The integral for $Q$ in Eq.(\ref{eq:Q}) would diverge at large $k$ if
there were no noise in the power spectrum $P(k)$, i.e., $P(k)=P_\mr{tot}(k)$. 
This happens when we use the dark matter density field from high precision 
$N$-body simulations to reconstruct the large-scale density field. 
The bias factor solves this problem.
The result can be improved in the future by including the higher-order perturbations. 
Nevertheless, the reconstruction works well even at this leading order, with 
the cross-correlation coefficient larger than 0.9 until $k\gtrsim0.1\mr{Mpc}/h$.
The reconstruction should work even better at higher redshifts since 
Eq. (\ref{eq26}) holds better when  
the small-scale density fluctuations undergo less nonlinear evolutions. 

The $\hat{\gamma}_1$ and $\hat{\gamma}_2$ estimators are optimal under 
the Gaussian assumption, while this is not the
real case. The non-Gaussian density field is treated as a Gaussian random
field throughout the reconstruction. So we need to Gaussianize the non-Gaussian
density field or the correlation would be rather weak.
It would be worthwhile to investigate the 
optimal estimators for non-Gaussian sources, which can be constructed from $N$-body
simulations.  

There is still one more limitation in the reconstruction.
The $\hat{\gamma}_1$ and $\hat{\gamma}_2$ estimators are unbiased
and of minimum variance in the long-wavelength limit, providing optimal tidal
shear reconstructions.
However, at larger wave numbers, the tidal shear field would be underestimated
by a scale-dependent multiplicative bias factor which needs to be corrected 
\cite{2008:lu,2012bucher}.
In this paper, we find the estimator derived in the long-wavelength limit
still works well if we only use the reconstruction modes with scale larger than
the smoothing scale.
However, when there is an overlap between these two scales, the reconstruction 
result would degrade. A more sophisticated method is needed in that case.   

In this paper, we use the dark matter density field directly to reconstruct the 
large-scale density field. However, the actual tracers would be galaxies, which
are located inside dark matter halos which are in turn distributed in the 
underlying dark matter density field. Due to the discreteness of the 
galaxy (halo) field, the tidal reconstruction may not be as good as that of the 
dark matter field. The scale-dependent galaxy (halo) bias may also complicate 
the reconstruction procedure. Further, the reconstruction will be affected by
the tidal galaxy (halo) bias. These effects can be quantified using the halo 
density fields from $N$-body simulations. 
We plan to study these in the future.

The reconstruction method developed in this paper has great potential applications.
The reconstructed field $\kappa$ gives the distribution of dark matter
on large scales. By cross correlating with the original galaxy (halo) field,
we can measure the logarithmic growth rate without sample variance 
\cite{2009mcdonald}. This can potentially provide high precision measurements 
of neutrino masses and tests of gravity. The reconstructed field $\kappa$ is also 
important for measuring the dark matter-neutrino cross-correlation dipole 
\cite{2014PhRvL.113m1301Z}. 
Moreover, it is also important in the case where 
the measurement of long-wavelength modes is missing or has large noise. 
In the 21cm intensity mapping survey \cite{2008PhRvL.100i1303C}, modes with 
small radial wave numbers are seriously contaminated by foreground radiation 
and are often subtracted in data analysis.
This makes it difficult or even impossible to do cross correlations with surveys 
which only probe the long-wavelength radial modes, including weak lensing, photo-$z$ 
galaxies, integrated Sachs-Wolf effect and kinetic Sunyaev-Zel'dovich effect.
Since cosmic tidal reconstruction can recover the long-wavelength radial modes, it 
enables the cross correlations of the 21cm intensity survey with other cosmic probes.
For 21cm intensity mapping experiments such as Tianlai 
\cite{2012IJMPS..12..256C,2015ApJ...798...40X} and CHIME \cite{2014SPIE.9145E..22B},
the cosmic tidal reconstruction method can be very useful.

%
%

\section*{acknowledgements}
The simulations were performed on the BGQ supercomputer at the SciNet HPC
Consortium. SciNet is funded by the Canada Foundation for Innovation under 
the auspices of Compute Canada, the Government of Ontario, the Ontario 
Research Fund Research Excellence, and the University of Toronto.
We acknowledge the support of the Chinese MoST 863 program under Grant 
No. 2012AA121701, the CAS Science Strategic Priority Research Program 
XDB09000000, the NSFC under Grants No. 11373030, No. 11403071, and No. 11473032,
IAS at Tsinghua University, CHEP at Peking University, and NSERC. 
Research at the Perimeter Institute is supported by the Government of Canada
through Industry Canada and by the Province of Ontario through the Ministry of
Research $\&$ Innovation.

\bibliographystyle{apsrev}
\bibliography{tide}
\end{document}